\documentclass[aps,pra,showpacs,twocolumn,amsmath,amssymb]{revtex4}

\usepackage[english]{babel}
\usepackage{latexsym}
\usepackage{graphics}
\usepackage{epsfig}
\usepackage{color}

\def\be{\begin{equation}}
\def\ee{\end{equation}}
\def\bea{\begin{eqnarray}}
\def\eea{\end{eqnarray}}
\def\bi{\begin{itemize}}
\def\ei{\end{itemize}}
\def\bin{\begin{enumerate}}
\def\ein{\end{enumerate}}
\def\la{\langle}
\def\ra{\rangle}

\usepackage{graphicx}
\begin{document}
\title{Dark soliton in a disorder potential}

\author{Ma\l{}gorzata Mochol}
\affiliation{
Instytut Fizyki imienia Mariana Smoluchowskiego and
Mark Kac Complex Systems Research Center, 
Uniwersytet Jagiello\'nski, ulica Reymonta 4, PL-30-059 Krak\'ow, Poland}

\author{Marcin P\l{}odzie\'n}
\affiliation{
Instytut Fizyki imienia Mariana Smoluchowskiego and
Mark Kac Complex Systems Research Center, 
Uniwersytet Jagiello\'nski, ulica Reymonta 4, PL-30-059 Krak\'ow, Poland}

\author{Krzysztof Sacha}
\affiliation{
Instytut Fizyki imienia Mariana Smoluchowskiego and
Mark Kac Complex Systems Research Center, 
Uniwersytet Jagiello\'nski, ulica Reymonta 4, PL-30-059 Krak\'ow, Poland}

\date{\today}

\begin{abstract}
We consider a dark soliton in a Bose-Einstein condensate in the presence of 
a weak disorder potential. Deformation of the soliton shape is analyzed within 
the Bogoliubov approach and by employing an expansion in eigenstates 
of the P\"oschl-Teller Hamiltonian. Comparison of the results with
the numerical simulations indicates that the linear response analysis reveals a 
good agreement even if the strength of disorder is of the order
of the chemical potential of the system. In the second part of the paper we
concentrate on the quantum nature of the dark soliton and demonstrate that the
soliton may reveal Anderson localization in the presence of disorder. 
The Anderson localized soliton may decay due to quasi-particle excitations
induced by the disorder. However, we show that the corresponding lifetime 
is much longer than the condensate lifetime in a typical experiment.
\end{abstract}

\pacs{03.75.Lm,72.15.Rn,05.30.Jp}

\maketitle

\section{Introduction}

Ultra-cold atomic gases become a playground where complex systems 
of solid state physics, nonlinear quantum optics or even
cosmology can be efficiently simulated and investigated \cite{bloch2008,jaksch2005}. The level of experimental 
control and detection is unprecedented and allows one to build quantum 
simulators, i.e. experimentally controlled systems that are able to mimic other 
systems difficult to investigate directly \cite{buluta2009}. Since the first laboratory 
achievement of Bose-Einstein condensation the list of problems investigated 
experimentally and theoretically in ultra-cold atomic gases becomes very 
long and includes: superfluid phases of both bosonic and fermionic atomic 
species \cite{jin,varenna}, collective excitations like solitons and vortices \cite{burger1999,denschlag2000,khaykovich2002,strecker2002,matthews1999,madison2000,abo2001}, 
insulating phases of solid state physics \cite{jaksch1998,fisher1989,greiner2002}, transport properties and Anderson
localization effects \cite{billy2008,roati2008,kondov2011,jendrzejewski2011}, 
atomic systems in the presence of artificial gauge potentials \cite{lin2009,lin2009a,dalibard2010}.

Particularly interesting is the interplay between particle interactions and 
disorder phenomena in quantum many body systems. Anderson localization \cite{anderson1958}, which 
is essentially a single particle phenomenon, is vulnerable to particle 
interactions \cite{damski2003,lye2005,fort2005,clement2005,schulte2005,schulte2006}. In experiments with atomic gases 
the localization was observed when the interactions were practically turned off 
by employing Feshbach resonances or by reducing density of the atomic cloud \cite{billy2008,roati2008,kondov2011,jendrzejewski2011}. 

Particle interactions play a vital role in soliton formation.
Mean field description of atomic Bose-Einstein condensates reduces to the
Gross-Pitaevskii equation \cite{gpe,dalfovo1999,castinleshouches} which possesses a localized wave-packet solution (i.e. 
bright soliton) for attractive particle interactions and a solution with a phase 
flip (i.e. dark soliton) for repulsive interactions. Both kinds of the solitons
have been observed experimentally \cite{burger1999,denschlag2000,khaykovich2002,strecker2002}. 

While the Gross-Pitaevskii equation is a single
particle description with the interactions included in the mean field 
approximation one can anticipate quantum many body effects that go beyond
such a description. Center of mass of a bright soliton is a degree of freedom
which, when properly described quantum mechanically, allows for the analysis of
interesting phenomena like soliton scattering on a potential barrier which leads
to a superposition of macroscopically distinct objects \cite{weiss-castin} or quantum entanglement of
a pair of solitons  \cite{lewenmalomed2009}. 
It is also shown that in spite of the fact that the particle interactions are
present and are responsible for the bright soliton formation the center of mass of
the soliton
can reveal Anderson localization in a weak disorder 
potential \cite{sacha2009prl,sacha2009app}. One may raise a natural question if similar localization can be
also observed in the case of a dark soliton. In the bright soliton case a huge energy
gap for quasi-particle excitations guarantees that the shape of the soliton is
not perturbed by a weak disorder potential. Then, the only degree of freedom 
affected by the disorder is the center of mass and (as has been shown recently) it becomes
Anderson localized. 
In the dark soliton
case the situation is more delicate because there is practically  
no gap for quasi-particle
excitations. Consequently one may expect that coupling between the degree of
freedom that describes soliton position and the quasi-particle subsystem 
induced by the disorder can destroy Anderson localization of the dark soliton. 
In the present paper we address this problem and show that the dark soliton
localization can be observed experimentally because the lifetime of the 
localized states is sufficiently long.

The paper is organized as follows. In Sec.~\ref{secii} we analyze the deformation of a dark
soliton solution in the presence of a weak external potential within the
Bogoliubov approach and by reducing the description to the problem of 
the P\"oschl-Teller
potential. Such a classical description of the soliton allows us to conclude
that the deformation effect is very weak. It implies that 
in the quantum description of the soliton we may expect that 
coupling between the degree of freedom that describes soliton 
position and the quasi-particle subsystem can be neglected. It is confirmed in
Sec.~\ref{seciii} where we switch to the quantum description and 
calculate the lifetimes of Anderson localized dark solitons. We conclude in Sec.~\ref{concl}.

\section{Classical description}
\label{secii}

We consider $N_0$ bosonic atoms with repulsive interactions in a 1D box 
potential of length $L$ at zero temperature (for experimental realization of a box potential see \cite{raizen}). The mean field description assumes that all atoms occupy
the same single particle state $\phi_0$ which is a solution of the Gross-Pitaevskii equation (GPE) \cite{gpe}
\be
-\frac{\hbar^2}{2m}\partial_x^2\phi_0+g_0|\phi_0|^2\phi_0=\mu_0\phi_0,
\label{gpe0}
\ee 
where $g_0 = 2\hbar a\omega_\perp$, $\mu_0$
is chemical potential of the system, 
$a$ stands for the $s$-wave scattering length of the atoms and $\omega_\perp$ denotes the transverse harmonic confinement frequency.
By virtue of the nonlinear term in the GPE there exists a stationary dark 
soliton solution which, far from the boundary of the box potential, takes the form
\begin{equation}
\phi_0(x-q)=e^{-i\theta}\sqrt{\rho_0}\textrm{ tanh}\left(\frac{x-q}{\xi}\right),
\label{sol0}
\end{equation}
where $\theta$ is an arbitrary phase, $\rho_0$ stands for the atomic density away from the soliton position $q$ and $\xi=\hbar/\sqrt{mg_0\rho_0}$ is the so-called healing length.  The chemical potential of the system is $\mu_0=g_0\rho_0$. 
The description of a dark soliton restricted to the classical wave equation (\ref{gpe0}) is called the classical description. It is in contrast to the quantum description of Sec.~\ref{seciii} where, for example, the position of a dark soliton $q$ becomes a quantum degree of freedom. We consider a finite 1D system. However, in order to describe the system in a region away from the boundary of the box potential we may use analytical solutions corresponding to the infinite configuration space which, if necessary, can be modified to satisfy boundary conditions \cite{dziarmaga2004}, e.g. $\phi_0(x=0)=0$ and $\phi_0(x=L)=0$.

In the remaining part of the paper we adapt the following units for energy, length and time, respectively 
\bea
E_0&=&\mu_0, \cr
l_0&=&\xi, \cr
t_0&=&\frac{\hbar}{\mu_0}.
\label{units}
\eea

\subsection{Deformation of a dark soliton: Expansion in Bogoliubov modes}

We would like to describe the deformation of a dark soliton solution in the presence of a weak disorder potential. The disorder potentials we are interested in correspond to optical speckle potentials which are created experimentally by shining laser radiation on a so-called diffusive plate \cite{schulte2005,clement2006}. In the far field, light forms a random intensity pattern which is experienced by atoms as an external disorder potential. Diffraction effects are responsible for a finite correlation length of the speckle potentials. Interestingly properties of such a disorder can be easily engineered that allows, e.g., for preparation of a matter-wave analog of an optical random laser \cite{plodzien2011}. 

Bright soliton deformation in an optical speckle potential has been already considered in Ref.~\cite{cord2011}, see also \cite{gredeskul1992}. Dark soliton propagation and its radiation in the presence of randomly distributed Dirac-delta potentials has been considered in Ref.~\cite{bilasPRL05}, see also \cite{kivsharPR98,frantzeskakisJPA10}. In the case of repulsive particle interactions the deformation of a condensate ground state in an optical speckle potential has been also a subject of scientific publications \cite{giorgini1994,sanchez-palencia2006,gaul2009,gaul2011}. In the present paper we consider a problem of the repulsive interactions but we deal with a localized soliton structure, i.e. there is a phase flip of the condensate wavefunction at the position of a soliton.

In our considerations the dark soliton is placed in a bounded and weak external potential $V(x)$. To calculate a small perturbation of the solitonic wavefunction we start with the time--independent GPE which in our units (\ref{units}) takes the form
\begin{equation}		-\frac{1}{2}\partial^2_x\phi(x)+\frac{1}{\rho_0}|\phi(x)|^2\phi(x)+V(x)\phi(x)=\mu\phi(x), 
\label{stacGP1D}
\end{equation}
and substitute $\phi=\phi_0+\delta\phi$ and $\mu=\mu_0+\delta\mu=1+\delta\mu$ where $\delta\phi$ is a small perturbation of the soliton wavefunction (\ref{sol0}) and $\delta\mu$ is a small contribution to the chemical potential which allows us to correct 
a possible change of the total particle number due to the presence of $V(x)$. Keeping linear terms only we obtain the time--independent, non--homogeneous Bogoliubov--de Gennes equations
\begin{equation}
{\cal L}\left[ {\begin{array}{cc}\delta\phi   \\
\delta\phi^\ast   \\
\end{array} } \right]
= 
V\left[ {\begin{array}{cc}
-\phi_0   \\
\phi^\ast_0   \\
\end{array} } \right]
+
\delta\mu\left[ {\begin{array}{cc}
\phi_0   \\
-\phi^\ast_0   \\
\end{array} } \right], \label{BdG}
\end{equation}
where 
\begin{equation}
{\cal L}=\left( {\begin{array}{cc}
-\frac{1}{2}\partial^2_x+\frac{2}{\rho_0}|\phi_0|^2-1 & +\frac{1}{\rho_0}\phi^2_0   \\
-\frac{1}{\rho_0}\phi^{\ast 2}_0 & \frac{1}{2}\partial^2_x-\frac{2}{\rho_0}|\phi_0|^2+1   \\
\end{array} } \right). \label{operator_L}
\end{equation}
In order to solve Eqs.~(\ref{BdG}) we would like to expand the two component vector $(\delta\phi,\delta\phi^*)^T$ in a complete basis that consists of eigenvectors of the non-Hermitian operator $\cal L$. This basis has been published in Ref.~\cite{dziarmaga2004}, see also \cite{lewenstein,castin-dum}, and here we only present the results and comment on essential elements of the derivation.
Collecting the complete basis one has to be careful because the ${\cal L}$ operator is not diagonalizable \cite{lewenstein,castin-dum,dziarmaga2004,sacha2009prl}. There are right eigenstates of $\cal L$, i.e. ${\cal L}|\psi_k\ra=\epsilon_k|\psi_k\ra$, where $|\psi_k\ra=(|u_k\ra,|v_k\ra)^T$ and \cite{dziarmaga2004}
\bea
u_k(x)&=&\frac{e^{ikx}e^{-i\theta}}{4\sqrt{\pi}\epsilon^{3/2}_k}\left[\left(k^2+
2\epsilon_k\right)\left(\frac{k}{2}+i\textrm{tanh}(x-q)\right)+\right.\cr && 
 \left.+\frac{k}{\textrm{cosh}^2(x-q)}\right], \label{u} \\
v_k(x)&=&\frac{e^{ikx}e^{i\theta}}{4\sqrt{\pi}\epsilon^{3/2}_k}\left[\left(k^2-
2\epsilon_k\right)\left(\frac{k}{2}+i\textrm{tanh}(x-q)\right)+\right. \cr &&
\left.+\frac{k}{\textrm{cosh}^2(x-q)}\right], \label{v}
\eea
that correspond to the familiar Bogoliubov spectrum 
\begin{equation}
\epsilon_k=\frac{1}{2}\sqrt{4k^2+k^4}.
\label{bogspec}
\end{equation}
These eigenstates are called phonons and their adjoint vectors $|\psi^{ad}_k\ra=(|u_k\ra,-|v_k\ra)^T$ are also the left eigenmodes of $\cal L$. Due to symmetries of the ${\cal L}$ operator $|\tilde\psi_k\ra=(|\tilde u_k\ra,|\tilde v_k\ra)^T=(|v^*_k\ra,|u_k^*\ra)^T$ are the right eigenmodes corresponding to eigenvalues $\tilde\epsilon_k=-\epsilon_k$ and $|\tilde\psi^{ad}_k\ra=(-|v^*_k\ra,|u_k^*\ra)^T$ are their adjoint vectors.
In the infinite configuration space right eigenvectors and adjoint vectors fulfill $\la\psi_k^{ad}|\psi_{k'}\ra=\langle u_k|u_{k'}\rangle-\langle v_k|v_{k'}\rangle=\delta(k-k')$.
In a box potential the wavevector becomes quantized, i.e. $k_n=n\pi/L$ where $n=1,2,\ldots$, and there is a tiny energy gap for the phonon excitations. 
 
Phonons apart, there exist two zero-eigenvalue modes of ${\cal L}$
corresponding to broken gauge $U(1)$ and translational symmetries \cite{dziarmaga2004},
\begin{equation}
\left[ {\begin{array}{cc}
u_\theta   \\
v_\theta  \\
\end{array} } \right]
=
i\hspace{0.05cm}\frac{\partial}{\partial\theta}\left[ {\begin{array}{cc}
\phi_0   \\
\phi^\ast_0  \\
\end{array} } \right], \label{zero:theta}
\end{equation} 
\begin{equation}
\left[ {\begin{array}{cc}
u_q   \\
v_q  \\
\end{array} } \right]
=
i\hspace{0.05cm}\frac{\partial}{\partial q}\left[ {\begin{array}{cc}
\phi_0   \\
\phi^\ast_0  \\
\end{array} } \right], \label{zero:q}
\end{equation}
respectively. They appear as zero-eigenvalue vectors because a shift of the global phase $\theta$ or a change of the soliton position $q$ in solution (\ref{sol0}) does not lead to a change of the system energy. 
The zero modes fulfill $\langle u_q|u_\theta\rangle-\langle v_q|v_\theta\rangle=0$ and they are orthogonal to the modes adjoint to the phonons. Vectors adjoint to the zero modes are not left eigenvectors of the $\cal L$ and they can be found by solving \cite{castin-dum}
\begin{equation}
{\cal L}\left[ {\begin{array}{cc}
u^{ad}_{\theta,q}    \\
v^{ad}_{\theta,q}   \\
\end{array} } \right]
=
\frac{1}{M_{\theta,q}}\left[ {\begin{array}{cc}
u_{\theta,q}    \\
v_{\theta,q}   \\
\end{array} } \right], \label{inhomo:eq}
\end{equation}
where $M_{\theta,q}$ are determined by normalization conditions $\langle u^{ad}_{\theta,q}|\hspace{0.05cm}u_{\theta,q}\rangle-\langle v^{ad}_{\theta,q}|\hspace{0.05cm}v_{\theta,q}\rangle=1$. 
One gets \cite{dziarmaga2004}
	\bea
		\left[ {\begin{array}{cc}
 				u^{ad}_\theta   \\
 				v^{ad}_\theta  \\
 			\end{array} } \right]
 	&=&
 	\frac{\partial}{\partial N_0}\left[ {\begin{array}{cc}
 						\phi_0   \\
 						\phi^\ast_0  \\
 					\end{array} } \right]
 		-iR\left[ {\begin{array}{cc}
 				u_q   \\
 				v_q  \\
 			\end{array} } \right], \label{eq:theta_ad}
	\\
		\left[ {\begin{array}{cc}
				u^{ad}_q   \\
				v^{ad}_q  \\
 			\end{array} } \right]
 			&=&
 		-\frac{i}{4\sqrt{\rho_0}}\left[ {\begin{array}{cc}
 							e^{-i\theta}  \\
							-e^{i\theta}  \\
 						\end{array} } \right], \label{eq:q_ad}
	\eea
where $M_\theta=\rho_0(\partial N_0/\partial \rho_0)$, $M_q=-4\rho_0$ and $R=(2q-L)\rho_0/M_qM_\theta$. We have $\langle u^{ad}_q|u_\theta\rangle-\langle v^{ad}_q|v_\theta\rangle=0$ and
$\langle u^{ad}_\theta|u_q\rangle-\langle v^{ad}_\theta|v_q\rangle=0$.
A small contribution of the zero mode $(u_q,v_q)^T$ in (\ref{eq:theta_ad}) allows one to fulfill also $\langle u^{ad}_q|u^{ad}_\theta\rangle-\langle v^{ad}_q|v^{ad}_\theta\rangle=0$.

Now we have all vectors to build a complete basis and the deformation of the soliton can be expanded in that basis
	\begin{align} \label{rozklad}
		\left[ \begin{array}{cc}
 				\delta\phi   \\
 				\delta\phi^\ast   \\
 			\end{array}  \right]
 		=&
		\hspace{0.1cm}\Delta\theta\left[ \begin{array}{cc}
 						u_\theta   \\
 						v_\theta   \\
					   \end{array}  \right]
 		+
 		P_\theta\left[ \begin{array}{cc}
 				u^{ad}_\theta \\
 				v^{ad}_\theta   \\
			    \end{array}  \right]
 		+
 		\Delta q\left[ \begin{array}{cc}
 				u_q \\
 				v_q  \\
			    \end{array}  \right]
 		+
 		P_q\left[ \begin{array}{cc}
 				u^{ad}_q \\
 				v^{ad}_q   \\
			     \end{array}  \right]  \nonumber \\
 		&+
 		\sum_k \left( b_k\left[ \begin{array}{cc}
 						u_k   \\
 						v_k  \\
					\end{array}  \right]
 		+b^\ast_k\left[ \begin{array}{cc}
 					v^\ast_k   \\
 					u^\ast_k   \\
 				\end{array}  \right]\right).  		
	\end{align}
Substituting (\ref{rozklad}) into (\ref{BdG}) results in
\bea
V\left[ \begin{array}{cc}
-\phi_0   \\
\phi_0^*   \\
\end{array}  \right]+\delta\mu\left[ \begin{array}{cc}
\phi_0   \\
-\phi_0^*   \\
\end{array}  \right]=\frac{P_\theta}{M_\theta}\left[ \begin{array}{cc}
u_\theta   \\
v_\theta   \\
\end{array}  \right]+
\frac{P_q}{M_q}\left[ \begin{array}{cc}
u_q   \\
v_q   \\
\end{array}  \right] \cr
+  \sum_k \epsilon_k\left(b_k\left[ \begin{array}{cc}
u_k   \\
v_k   \\
\end{array}  \right]-b_k^*\left[ \begin{array}{cc}
v_k^*   \\
u_k^*   \\
\end{array}  \right] \right).
\label{project}
\eea
Projecting this equation onto the adjoint vectors we can obtain the expansion coefficients and a small correction to the chemical potential. 
Note that, there is no restriction for a choice of small deviation $\Delta\theta$ and $\Delta q$ which is due to the fact that these coefficients are related to the zero modes. 
However, while $\theta$ can be arbitrary, we will see that $q$ can not because the external potential breaks the translational symmetry but does not affect the $U(1)$ symmetry. Coefficient $\Delta\theta$ is related to a shift of the global phase of the soliton solution (\ref{sol0}) and without loss of generality we may choose $\Delta\theta=0$. For the coefficient $P_\theta$ we get
\bea
\frac{P_\theta}{M_\theta}&=&-2\la\partial_{N_0}\phi_0|V\phi_0\ra+\delta\mu \cr &&
-iR\left(\la u_q|V\phi_0\ra+\la v_q|V\phi_0^*\ra\right).
\label{ptheta}
\eea
The last term on the r.h.s. can be written in the following way
\bea
\la u_q|V\phi_0\ra+\la v_q|V\phi_0^*\ra\sim
\int_0^Ldx|\phi_0(x-q)|^2\partial_xV(x),
\label{force}
\eea
which represents a force acting on the soliton. A similar effective force appears in the approach of Ref.~\cite{pelinovskyZAMP08}. For the stationary solution it is obvious that the soliton position $q$ is chosen so that such a force is zero. Then, also an arbitrary shift of the soliton position should be zero, i.e. $\Delta q=0$ in (\ref{rozklad}). With such a choice of the soliton position the coefficient $P_\theta$ is 
directly related to a change of the total particle number which we assume to be zero. Hence, $P_\theta=dN=0$ and Eq.~(\ref{ptheta}) allows us to obtain the correction to the chemical potential
\bea
\delta\mu &=& 2\langle\partial_{N_0}\phi_0|V\phi_0\rangle \cr &=&
\frac{1}{L}\int_0^L{dy\left(\tanh y+y\;\textrm{sech}^2y\right)
		\tanh y V(y+q)}. \cr &&
\label{eq:delta_mu_B}
\eea
Projecting (\ref{project}) onto the adjoint mode (\ref{eq:q_ad}) results in
$P_q=0$ what can be expected because $P_q$ has an interpretation of the soliton momentum and for the stationary state it should be zero.

Thus, with a proper choice of the soliton position and a suitable correction of the chemical potential, all coefficients in (\ref{rozklad}) are zero except 
	\begin{equation}
		b_k=\frac{1}{\epsilon_k}\left[-\langle u_k|V\phi_0\rangle-\langle v_k|V\phi^\ast_0\rangle\hspace{0.05cm}\right], \label{eq:b_n}
	\end{equation}
which contain full information about soliton deformation in a weak external potential. Finally the stationary solitonic solution in the presence of a weak external potential reads
\be
\phi(x)=\phi_0(x)+\sum_k\left[b_ku_k(x)+b_k^*v_k^*(x)\right].
\label{fbog}
\ee

\subsection{Deformation of a dark soliton: Expansion in modes of  the P\"oschl--Teller potential}

The Bogoliubov approach is suitable in the description of the eigenmodes of collective or elementary excitations of a Bose-Einstein condensate \cite{dalfovo1999,castinleshouches}. However, if we are interested in the description of a stationary state of the GPE and if such a state can be represented as a real function one may introduce a substantial simplification. We will see that description of dark soliton deformation reduces to an expansion of a wavefunction perturbation in modes of the P\"oschl--Teller potential \cite{poschl-teller}.
Such an approach has been applied in the analysis of a bright soliton deformation in a weak disorder potential \cite{cord2011} and it turns out it is also suitable in the dark soliton case.

Let us start again with the stationary GPE but assume the solution we are looking for is a real function
\begin{equation} \left(-\frac{1}{2}\partial^2_x+\frac{1}{\rho_0}\phi^2-\mu+V(x)\right)\phi=0.
\end{equation}
Similarly to the Bogoliubov approach we introduce $\mu=\mu_0+\delta\mu=1+\delta\mu$ and $\phi=\phi_0+\delta\phi$ but assume that in (\ref{sol0}) the global phase $\theta=0$. Keeping linear terms only, rewriting $\phi^2_0(x-q)=\rho_0\tanh^2(x-q)=\rho_0\left[1-\cosh^{-2}(x-q)\right]$ and changing variable $x\rightarrow x+q$ we obtain
\begin{equation} 
  \left(H_0+2\right)\delta\phi=\delta\mu\phi_0-V\left(x+q\right)\phi_0, \label{H_0}
\end{equation}
where
\be
H_0=-\frac{1}{2}\partial^2_x-\frac{3}{\cosh^2(x)},
\ee
is the Hamiltonian for a particle in the P\"oschl--Teller potential \cite{poschl-teller}.
To compute $\delta\phi$ we need to invert the operator $H_0+2$. All the eigenstates of the Hamiltonian $H_0$ are known in the literature \cite{lekner2007}. There are two bound states
	\bea
	      \psi_0(x)&=&\frac{\sqrt{3}}{2}\textrm{sech}(x)^2 \label{eq:psi_0},\\
	      \psi_1(x)&=&\sqrt{\frac{3}{2}}\textrm{sech}(x)\tanh(x), \label{eq:psi_1}
	\eea
with eigenenergies $E_0=-2$ and $E_1=-\frac{1}{2}$ respectively and scattering states
	\begin{equation}      \psi_k(x)=\frac{e^{ikx}}{(2\pi)^{1/2}}\frac{k^2-2+3\textrm{sech}(x)^2+3ik\tanh(x)}{[(1+k^2)(4+k^2)]^{1/2}}, \label{eq:psi_k}
	\end{equation}
with $E_k=\frac{k^2}{2}$, $k\in\mathbb{R}$. 
We can therefore expand the deformation $\delta\phi$ over orthonormal basis of eigenfunctions
	\begin{equation} \delta\phi=\alpha_0\;\psi_0+\alpha_1\;\psi_1+\int{dk\;\alpha_k\;\psi_k(x)}, \label{rozklad_alfa}
	\end{equation}
and compute coefficients $\alpha_j$ by projecting Eq.~(\ref{H_0}) on the proper eigenmodes
	\begin{equation}\label{alfa}
	  (E_j+2)\alpha_j=\int{dx\;\psi^\ast_j(x)[\delta\mu\phi_0-V(x+q)\phi_0]}. 
	\end{equation}
The wavefunction (\ref{eq:psi_0}) is a zero mode, i.e. $(H_0+2)\psi_0=0$. Thus, Eq.~(\ref{H_0}) can be solved provided the projection of its r.h.s on the zero mode vanishes. Therefore we require
\bea
-\la\psi_0|V\phi_0\ra+\delta\mu\la\psi_0|\phi_0\ra&=&-\la\psi_0|V\phi_0\ra \sim \la\partial_x\phi_0|V\phi_0\ra \cr
&=&-\int\limits_0^L{dx\;\phi^2_0(x-q)\partial_xV(x)}\cr &=&0. 
\label{require}
\eea
In (\ref{require}) we have taken advantage of $\la\psi_0|\phi_0\ra=0$ and the fact that the zero mode is also the translation mode of the dark soliton, i.e. $\psi_0\sim \partial_x\phi_0$. Condition (\ref{require}) implies that the soliton position $q$ should be chosen so that the force acting on it is zero, compare with (\ref{force}). From Eq.~(\ref{alfa}) we do not get any restriction for the value of $\alpha_0$. However, because $\alpha_0\psi_0$ has an interpretation of a shift of the soliton position we should choose $\alpha_0=0$ if we are interested in a stationary solution.

In order to solve Eq.~(\ref{H_0}) we have to invert the operator $H_0+2$ in the Hilbert space with the zero mode excluded which is simple because all eigenfunctions of $H_0$ are known. That is 
\begin{equation}
  \delta\phi(x)=\int{dy\;K(x,y)\;[\delta\mu\phi_0-V(y+q)\phi_0]}, \label{jadro}
\end{equation}
where the symmetric kernel $K(x,y)$ reads
\bea
K(x,y)&=&\frac{2}{3}\psi_1(x)\psi^\ast_1(y)+2\int{\frac{\psi_k(x)\psi^\ast_k(y)}{4+k^2}}
\cr &=&
-\frac{1}{16}\textrm{sech}^2(x)\textrm{sech}^2(y)\times \cr
&&\left\{\textrm{sh}^22x+\textrm{sh}^22y+4\textrm{ch}2x+4\textrm{ch}2y\right. \cr
&&\left.-3-\left(\textrm{ch}2x+\textrm{ch}2y+3\right)|\textrm{sh}2x-\textrm{sh}2y|\right. 
\cr		&&\left.-4\textrm{sh}|x-y|\textrm{sh}x\hspace{0.05cm}\textrm{sh}y-6|x-y|\right\}. \label{Cord}
\eea
In the Bogoliubov approach, even if we are restricted to the phonon subspace, the $\cal L$ operator possesses a zero eigenvalue if the configuration space is infinite, see (\ref{bogspec}). Therefore it is not straightforward to obtain a simple form of the relevant kernel because integration over phonon subspace should actually be substituted by summation over discrete values of the wavevector.

Finally we have to determine the correction to the chemical potential. In the Bogoliubov approach there is a specific mode responsible for a change of the total particle number which is orthogonal to all other Bogoliubov modes and to keep the particle number constant it is sufficient to ensure that the wavefunction perturbation does not have any contribution from this mode. In the present approach the chemical potential has to be determined by the normalization condition $\la\phi|\phi\ra=N_0+{\cal O}(\delta\phi^2)$ that implies $\la\phi_0|\delta\phi\ra=0$ and thus
\bea
\delta\mu&=&\frac{\int{dxdy\;\phi_0(x)K(x,y)V(y+q)\phi_0(y)}}{\int{dxdy\;\phi_0(x)K(x,y)\phi_0(y)}} \cr
&=&\frac{1}{L}\int{dy\left(\tanh y+y\;\textrm{sech}^2y\right)\tanh y V(q+y)}, 
\cr &&
\eea
which is the same expression as in the Bogoliubov approach, compare with (\ref{eq:delta_mu_B}). 

The final expression for a stationary solitonic solution in the presence of a weak external potential reads
\bea
\phi(x)&=&\phi_0(x)-\int{dy\;K(x,y)V(y+q)\phi_0(y)} \cr
&& +\delta\mu \;\frac{\partial \phi_0(x)|_{\mu_0}}{\partial\mu_0},
\label{finalpt}
\eea
where we have made use of $\int{dyK(x,y)\phi_0(y)}=\frac{\partial \phi_0(x)|_{\mu_0}}{\partial\mu_0}$ with $\phi_0|_{\mu_0}\equiv\phi_0$ being a wave function 
with fixed chemical potential $\mu_0=1$.

In the bright soliton case \cite{cord2011} the eigenfunction $\psi_1$ (not $\psi_0$ as in the dark soliton problem) turns out to be a translation mode of the system and also the $H_0+2$ operator is substituted by $H_0+1/2$. Consequently, one obtains a different expression for the symmetric kernel which is expected because the bright soliton represents a localized wavepacket while a dark soliton is a phase flip at soliton position with otherwise uniform density.  

\subsection{Comparison with numerical calculations}

We are going to compare the perturbative approaches introduced in the previous subsections with results of numerical calculations in the case of a dark soliton in a weak optical speckle potential. However, first we consider a simple harmonic potential 
\be
V(x)=-V_0\cos(k_0x),
\label{vcos}
\ee
which on one hand is a generic example because any potential can be expanded in a Fourier basis and on the other hand allows us to obtain simple expressions
for expansion coefficients. 

In the Bogoliubov approach Eq.~(\ref{eq:b_n}), for the potential (\ref{vcos}), one gets
\bea
b_k&=&-iV_0\sqrt{\rho_0}\frac{\sqrt{\pi}}{2k(4+k^2)\sqrt{\epsilon_k}}\times
\cr &&
\left\{
k_0^2\left({\rm csch}\frac{\pi(k-k_0)}{2}+{\rm csch}\frac{\pi(k+k_0)}{2}\right)\right.
\cr &&
\left.
+4k\;\left[\delta(k-k_0)+\delta(k+k_0)\right]\;
\right\}.
\label{bcos}
\eea
This coefficient contains Dirac delta functions which tell us that the condensate density will be modulated with a period of $2\pi/k_0$. At small $k$ the coefficient diverges
\be
b_k\sim \frac{1}{\sqrt{|k|}}+{\cal O}(|k|^{3/2}),
\ee
but the contribution to the soliton wavefunction from long wavelength phonons is finite
\be
b_ku_k+b_k^*v_k^*\sim {\rm const}+{\cal O}(k^2).
\ee
Thus, in order to calculate soliton deformation we may assume infinite configuration space because the singularity at $k\rightarrow 0$ is actually not dangerous.
The correction to the chemical potential $|\delta\mu|<\frac{4V_0}{k_0L}$, see Eq.~(\ref{eq:delta_mu_B}), disappears in the limit of $k_0L\rightarrow\infty$ which is a consequence of the self-averaging property of the potential (\ref{vcos}), see \cite{sanchez-palencia2006}.

\begin{figure}
\centering
\includegraphics*[scale=0.33]{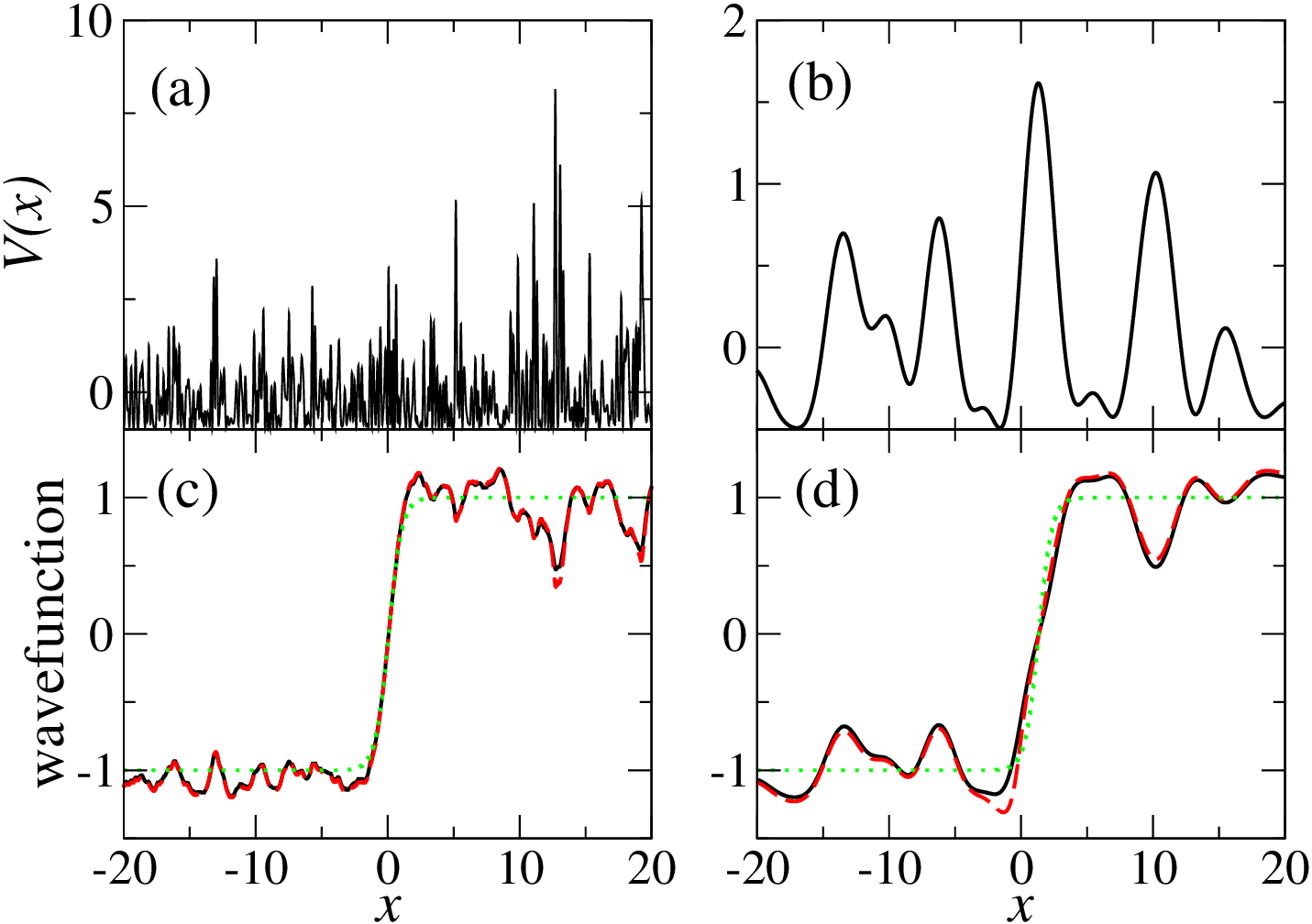}
\caption{(Color online) 
In panel (a) we show an example of the optical speckle potential with the correlation length $\sigma_R=0.05$ and for $V_0=1$ while in panel (c) we present the corresponding exact solution of the Gross-Pitaevskii equation obtained numerically (solid black line) and within the perturbation approach (red dashed line), see Eq.(\ref{finalpt}) (Eq.(\ref{fbog}) leads to the same results). In panels (b) and (d) we show the same as in (a) and (c) but for $\sigma_R=1$ and $V_0=0.5$. Green dotted lines in (c) and (d) correspond to unperturbed soliton wavefunctions (\ref{sol0}).}
\label{speckle}
\end{figure}

In the approach that involves modes of the P\"oschl-Teller potential Eq.~(\ref{alfa}) yields 
\bea
\alpha_k&=&iV_0\sqrt{\rho_0}\frac{\sqrt{\pi}}{\sqrt{8}(1+k^2)^{1/2}(4+k^2)^{3/2}}\times \cr &&\left\{
(k^2-3k^2_0+4)\left({\rm csch}\frac{\pi(k-k_0)}{2}+{\rm csch}\frac{\pi(k+k_0)}{2}\right)\right. \cr
&&\left.-12k\;[\delta(k+k_0)+\delta(k-k_0)]\;\right\}, 
\label{acos}
\eea
for scattering modes and 
\be
\alpha_1=
V_0\sqrt{\frac{2\rho_0}{3}}\frac{\pi}{2}(1-k_0^2)\;\textrm{sech}\left(\frac{k_0\pi}{2}\right),
\ee
for the bound state $\psi_1$,
where we have assumed that the system is infinite and thus the correction $\delta\mu=0$. There are Dirac delta functions present in (\ref{acos}) similarly as in the Bogoliubov coefficients but contrary to the Bogoliubov case the coefficients $\alpha_k$ possess no singularity at $k=0$. 

Note that the coefficients (\ref{bcos}) and (\ref{acos}) are proportional to $V_0$. Thus, for $V_0\ll 1$ (i.e. for the potential strength much smaller than the chemical potential of the system) the perturbation of the condensate wavefunction is certainly negligible. 
Deformation of the soliton shape depends on the relation of $k_0$ to the inverse of the healing length. For $k_0\ll1$, i.e. if the potential changes on a scale much larger than the healing length of the system, we may employ the Thomas-Fermi profile \cite{gpe} and approximate the condensate wavefunction by $\phi(x)\approx\sqrt{\rho_0[1-V(x)]}\tanh(x/\tilde\xi)$ where $\tilde\xi=1/\sqrt{1+V_0}$ is a local healing length around the soliton position. Thus, for $k_0\ll1$, the soliton shape is still given by the hyperbolic tangent function but its size is smaller. In the other limit, i.e. for $k_0\gg1$, the condensate wavefunction reveals harmonic oscillations superposed on the hyperbolic tangent function. Indeed, for $k_0\gg 1$, Eq.~(\ref{acos}) implies that the dominant contribution comes from
$k\approx k_0$ which behaves like $V_0/k_0^{2}$.
Thus, even if we increase $V_0$ but choose $k_0\gg 1$ the wavefunction perturbation is still negligible due to smoothing effects. 
Finally we would like to stress that plotting $\phi$ obtained in the Bogoliubov approach and with the help of the P\"oschl-Teller modes one gets exactly the same results. 

For comparison with numerics we have chosen the case of an optical speckle potential. Such a potential is characterized by: zero mean value $\overline{V(x)}=0$, where the overbar denotes an ensemble average over disorder realizations, standard deviation
$V_0=\left[\overline{V(x)^2}\right]^{1/2}$ and autocorrelation function $\overline{V(x')V(x'+x)}=V_0^2\frac{\sin^2(x/\sigma_R)}{(x/\sigma_R)^2}$ where $\sigma_R$ is the correlation length of the disorder. In Fig.~\ref{speckle} we show examples of the solitonic solutions in the presence of an optical speckle potential for a correlation length much smaller ($\sigma_R=0.05$)
and comparable ($\sigma_R=1$) to the healing length and for $V_0=1$ and $V_0=0.5$  respectively, obtained within the perturbation approaches and by a numerical solution of the GPE. The agreement is surprisingly good even though the strength of the disorder is of the order of the chemical potential. For $\sigma_R=0.05$ the disorder changes rapidly and its effect on the condensate is significantly smaller than for $\sigma_R=1$. 

We have considered a dark soliton in a box potential and analyzed its deformation due to the presence of a weak disorder potential. Our results can also be applied to the system in the presence of, e.g., a shallow harmonic trap. Indeed, if we are interested in the deformation of the condensate wavefunction in the vicinity of the trap center and if the change of the harmonic potential energy on a scale of the soliton size is much smaller than the disorder strength, i.e. $\omega^2\ll V_0$ where $\omega$ is harmonic trap frequency, the effect of the presence of the trap on the soliton deformation can be neglected. 

\section{Quantum description}
\label{seciii}

In the previous section, in order to describe ultra-cold atoms, we have applied the mean field approximation where it is assumed that a many body system is in a state where all atoms occupy the same single particle wavefunction. Solution of the GPE is an optimal choice for such a single particle wavefunction. Then, a stationary dark soliton appears as a solution of the classical wave equation and its position is given by a real number $q$. In the present section we will take into account situations when particles do not necessary occupy the same single particle state. It turns out that the problem can be described within the quantum version of the Bogoliubov approach where, e.g., $q$ becomes a quantum mechanical operator $\hat q$ and the soliton position is described by a probability distribution.
This approach has rather a semiclassical nature. The full quantum analysis would involve the N-body problem as has been done in Refs.~\cite{lai_haus1,lai_haus2} for the bright solitons in optical fibres, see also Ref.~\cite{castinleshouches}.

\subsection{Effective Hamiltonian}

An effective Hamiltonian that describes a bright soliton in the presence of a weak external potential has been introduced in Ref.~\cite{sacha2009app}. It is based on the Dziarmaga idea of how to describe non-perturbatively degrees of freedom corresponding to Bogoliubov zero modes \cite{dziarmaga2004}. Derivation of the effective Hamiltonian in the dark soliton case follows the same reasoning and therefore we will present key elements only.

In the previous section once the global phase $\theta$ of the wavefunction (\ref{sol0}) and the soliton position $q$ have been chosen no deviations of them have been considered. Small deviations can be described with the help of the zero modes, compare (\ref{zero:theta})-(\ref{zero:q}) and (\ref{rozklad}), while large deviations need modifications of the description. The expansion of the wavefunction perturbation around a given value of the soliton position $q$, see (\ref{rozklad}), is actually not necessary because one may treat $q$ as a dynamical variable and the same is true for $\theta$ \cite{dziarmaga2004}. This way we obtain 
\bea \label{rozklad_nonpert}
    \left[ \begin{array}{cc}
 				\phi   \\
 				\phi^\ast   \\
 			\end{array}  \right]
 		&=&
		\left[ \begin{array}{cc}
 						\phi_0   \\
 						\phi^\ast_0  \\
					   \end{array}  \right]
 		+
 		P_\theta\left[ \begin{array}{cc}
 				u^{ad}_\theta \\
 				v^{ad}_\theta   \\
			    \end{array}  \right]
 		+
 		P_q\left[ \begin{array}{cc}
 				u^{ad}_q \\
 				v^{ad}_q   \\
			     \end{array}  \right]  \cr
 		&&+
 		\sum_k \left( b_k\left[ \begin{array}{cc}
 						u_k   \\
 						v_k  \\
					\end{array}  \right]
 		+b^\ast_k\left[ \begin{array}{cc}
 					v^\ast_k   \\
 					u^\ast_k   \\
 				\end{array}  \right]\right).  		
\eea  
In (\ref{rozklad_nonpert}) all modes depend on $q$ and $\theta$ and can follow large changes of the soliton position and global phase.
Substituting (\ref{rozklad_nonpert}) into the energy functional
\be
H=\int{dx\left[\frac{1}{2}|\partial_x \phi|^2+V|\phi|^2+\frac{1}{2\rho_0}|\phi|^4-\mu|\phi|^2\right]}, \label{energy}
\ee
leads to
\bea \label{hamiltonian_class}
   H&=&-\frac{P_q^2}{2|M_q|}+\int{dx V(x)|\phi_0(x-q)|^2} \cr &&
   +\frac{P_\theta^2}{2M_\theta}+2P_\theta\la u^{ad}_\theta|V\phi_0\ra\cr
		&&+\sum_k [\epsilon_k b^\ast_k b_k +s_k(b_k+b_k^\ast)], 	
\eea
with
\be
 s_k = \la u_k|V\phi_0\ra+\left<v_k|V\phi_0^\ast\right>,
\ee
where only terms of the order ${\cal O}(P^2,b^2,PV,bV)$ are kept. 
Note that Eq.~(\ref{hamiltonian_class}) is the Hamiltonian formulation  of the classical perturbation theory applied in Sec.IIA. That is, fixed points of the Hamilton equations generated by (\ref{hamiltonian_class}) \cite{sacha2009app} correspond to stationary solutions analyzed in Sec.IIA. We know from Sec.II that such stationary solutions are in a very good agreement with the exact numerical calculations for the disorder strength even of the order of the chemical potential of the system, i.e. for $V_0\approx 1$. 

In the so-called second quantization formalism the quantum many body Hamiltonian corresponds to (\ref{energy}) where the wavefunction $\phi$ is substituted by a bosonic field operator $\hat\phi$. Then, also the expansion coefficients in (\ref{rozklad_nonpert}) become operators
\bea
 \hat{P}_q &=& -i\partial_q, \\
 \hat{P}_\theta &=& \hat{N} - N_0 = -i\partial_\theta,
\eea
and fulfill commutation relations
\bea
 [\hat{q},\hat{P}_q]&=&i, \cr
 [\hat{\theta},\hat{P}_\theta]&=&i,\cr
 [\hat{b}_k,\hat{b}^\dagger_{k'}] &=& \delta_{kk'}.
\eea
Energy functional (\ref{hamiltonian_class}) does not depend on $\theta$, thus, in the quantum description $[\hat{P}_\theta,\hat{H}] = 0$ and we may restrict it to the Hilbert subspace with exactly $N_0$ particles, i.e. for any state in this subspace $\hat{P}_\theta|\psi\ra=0$, and the quantum effective Hamiltonian reduces to
\begin{equation}
 \hat{H} = \hat{H}_q + \hat{H}_B + \hat{H}_1,
\end{equation}
where
\bea
 \hat{H}_q & =& -\frac{\hat{P}_q^2}{2 |M_q|} + \int dx V(x)|\phi_0(x-q)|^2 \cr
           &=&-\left(\frac{\hat{P}_q^2}{2 |M_q|}+\frac{|M_q|}{4}\int dx\frac{V(x)}{\cosh^2(x-q)}\right),
\label{hq}	     \\
 \hat{H}_B & =& \sum_k \epsilon_k \hat{b}^\dagger_k\hat{b}_k, \\
 \hat{H}_1 & =& \sum_k s_k(\hat{b}_k+\hat{b}^\dagger_k).
 \label{h1}
\eea
The Hamiltonian $\hat{H}_q$ describes soliton motion in an effective potential which turns out to be a convolution of the original potential with the density $|\phi_0|^2$. Owning to $|\phi_0|^2=\rho_0\tanh^2(x-q)=\frac{|M_q|}{4}[1-\cosh^2(x-q)]$ and $\overline{V(x)}=0$, $\hat H_q$ becomes similar to the corresponding Hamiltonian for a bright soliton in a weak external potential \cite{sacha2009prl}.
The term $\hat{H}_B$ describes quasi-particle subsystem (phonons) and $\hat{H}_1$ is the part of the Hamiltonian that couples soliton position degree of freedom with phonons. In the classical description (Sec.~\ref{secii}) 
such a coupling is responsible for the deformation of the stationary condensate wavefunction. In the following we will not look for eigenstates of the total Hamiltonian $\hat H$ but rather consider eigenstates of $\hat H_q$ and calculate the lifetime of the system prepared in these eigenstates due to the coupling with the quasi-particle subsystem induced by $\hat H_1$. 

We would like to emphasize striking differences between the classical description and the present quantum description. They are most apparent in the absence of an external potential. Indeed, for $V(x)=0$, the soliton position in the classical decription can be chosen arbitrary but it is well defined. In the quantum approach the Hamiltonian $\hat H_q$ tells us that eigenstates of the system correspond to eigenstates of the momentum operator $\hat P_q$ and the corresponding probability distributions for the soliton position are totally delocalized. Thus, similarly as in the bright soliton case \cite{castinleshouches}, dramatic condensate fragmentations are predicted in the quantum approach of a dark soliton \cite{dark}. 

\subsection{Anderson localization of a dark soliton}

The final form of the Hamiltonian $\hat H_q$ is similar to the Hamiltonian for  the center of mass of a bright soliton in a weak external potential \cite{sacha2009app}. The effective mass $|M_q|$ in (\ref{hq}) is equal to two times the number of particles missing in a dark soliton notch while in the bright soliton case it is given by a total number of particles in a system. A bright soliton is the ground state solution of the GPE and excitation of its center of mass increases the energy of the system. A dark soliton corresponds to a collectively excited system and in order to decrease the system energy one has to, e.g., accelerate the soliton. Indeed, excitations of the soliton position degree of freedom actually decrease the system energy due to the minus sign in front of the expression (\ref{hq}).

\begin{figure}
\centering
\includegraphics*[scale=0.33]{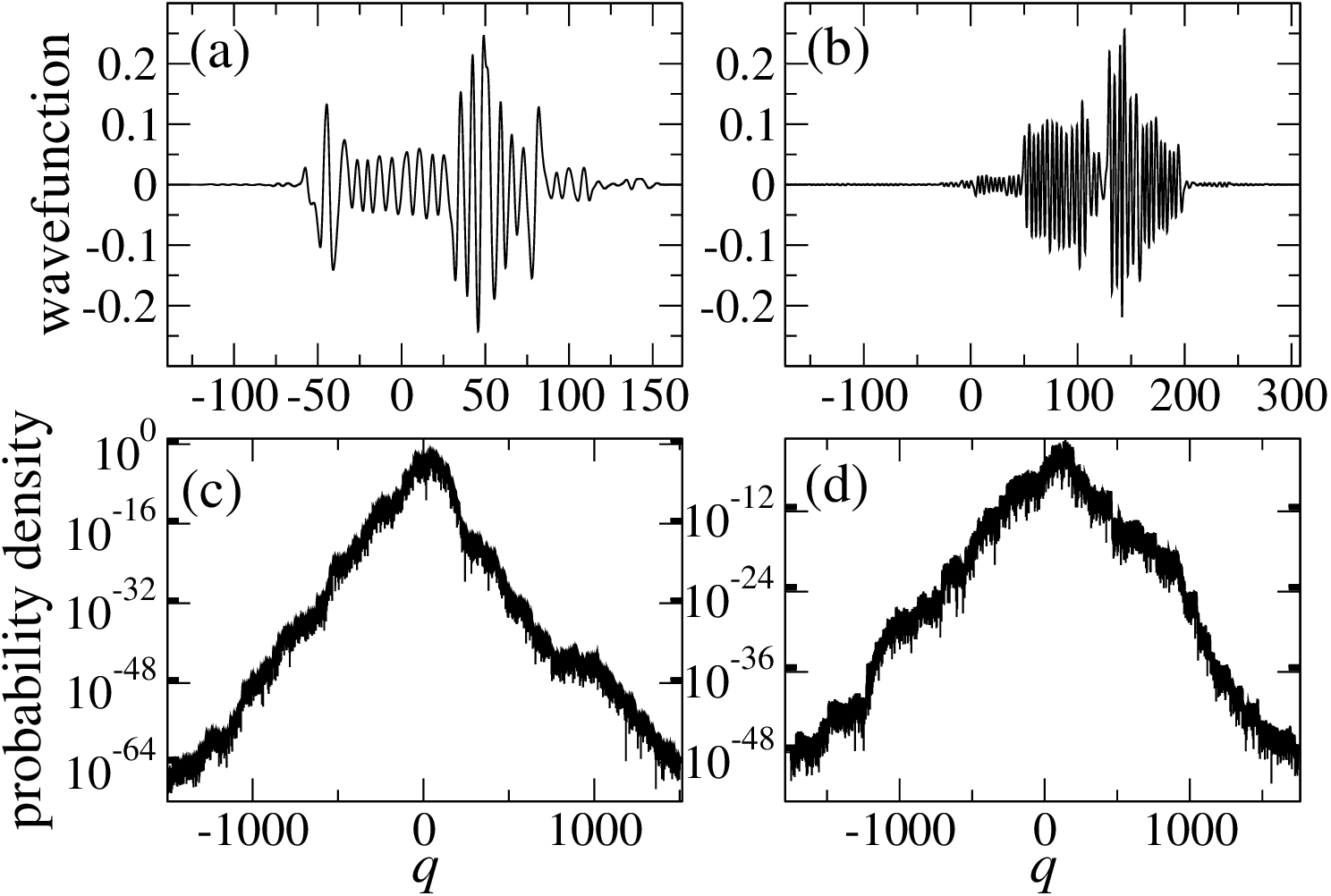}
\caption{
In top panels we show examples of eigenstates of the effective Hamiltonian $\hat H_q$, see (\ref{hq}), while in bottom panels the corresponding probability densities in log scale.
The correlation length of the speckle potential $\sigma_R=0.28$ and the strength $V_0=7\times 10^{-5}$ (left panels) and $V_0=1.4\times 10^{-4}$ (right panels). The eigenstates correspond to the eigenvalue $E_n=-3.03\times 10^{-3}$ (left panels) and $E_n=-8.58\times 10^{-3}$ (right panels) and reveal the localization length $l_{loc}=10.5$ and $l_{loc}=15.7$, respectively. 
}
\label{ander}
\end{figure}

It has been shown that in the presence of a weak disorder potential the center of mass of a bright soliton reveals Anderson localization \cite{sacha2009prl}. The same phenomenon can be expected for dark solitons. For $V(x)$ being an optical speckle potential with the correlation length $\sigma_R$ smaller than the healing length of the system we obtain the effective potential in (\ref{hq}) where the healing length plays a role of the effective correlation length. The generic properties of the Anderson localization in 1D allow us to expect that all eigenstates of the Hamiltonian $\hat H_q$ are exponentially localized, i.e., have a shape with the overall envelope \cite{lifshitz,tiggelen}
\begin{equation}
 |\psi_n(q)|^2 \propto \exp\left(-\frac{|q-q_0|}{l_{loc}}\right),
\end{equation}
where $\hat H_q\psi_n(q)=E_n\psi_n(q)$, $q_0$ is the mean position of the soliton and $l_{loc}=l_{loc}(E_n)$ is the localization length.
Indeed, in Fig.~\ref{ander} we present examples of the Anderson localized eigenstates for two values of the standard deviation of the speckle potential $V_0$.
The parameters we have chosen correspond to: $N_0=10^5$  $^{87}$Rb atoms in a quasi-1D box potential of $L=3550$ (3.37~mm) with the harmonic potential of $\omega_\perp=2\pi\times 370$~Hz in the transverse directions; the correlation length of the speckle potential $\sigma_R=0.28$ (0.27~$\mu$m)\cite{billy2008}. The energy units (\ref{units}) are the following: $E_0/\hbar=789$~Hz, $l_0=0.96$~$\mu$m and $t_0=1.27$~ms. 

In order to obtain predictions for Anderson localization of solitons we have neglected coupling of the soliton position degree of freedom to the quasi-particle subsystem. In the bright soliton case such an approximation is justified because there is a huge energy gap for quasi-particle excitations and if the strength of the potential is much smaller than the chemical potential of the system, corrections to the effective Hamiltonian $\hat H_q$ are negligible \cite{sacha2009app}. In the dark soliton case there is practically no energy gap for phonon excitations, i.e. minimal $\epsilon_k$, see (\ref{bogspec}), corresponds to $k=\pi/L$ which tends to zero for a large system. Moreover, a dark soliton is a collectively excited state which may decay to lower energy states by {\it emission} of phonons. If the strength of the disorder potential $V_0\ll 1$ we know from the classical analysis that the shape of a stationary dark solution of the GPE is negligibly deformed by the external potential. In the quantum description we may thus expect that the lifetime of Anderson localized eigenstates is sufficiently long and allows for experimental observations of the localization effects. 

\begin{figure}
\centering
\includegraphics*[scale=0.33]{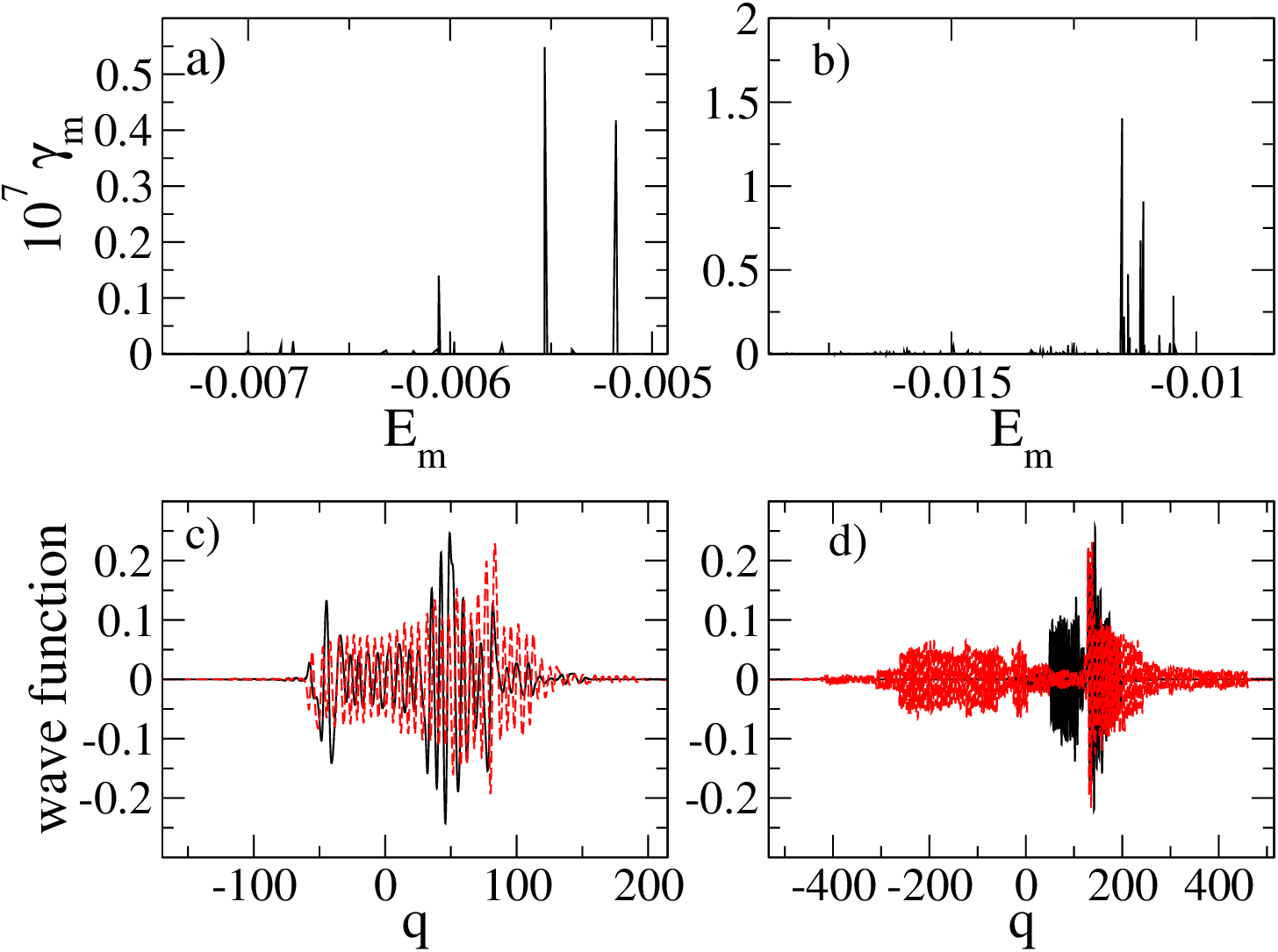}
\caption{(Color online) 
Top panels: contributions $\gamma_m$, Eq.~(\ref{maleg}), to the total decay rate $\Gamma$ as a functions of energy $E_m$. Bottom panels: initial states $\psi_n(q)$ of the system (solid black lines) and eigenstates $\psi_m(q)$ (dashed red lines) corresponding to the most probable decay channels, i.e. maximal values of $\gamma_m$. Parameters chosen in left (right) panels are the same as in the corresponding panels in Fig.~\ref{ander}.
}
\label{gam}
\end{figure}

Suppose we choose an initial state $|\Psi\ra$ of the $N_0$-particle system where the soliton position is described by an eigenstate $\psi_n(q)$ of the Hamiltonian $\hat H_q$ corresponding to an eigenvalue $E_n$ and there is no phonon excitation, i.e. we deal with the quasi-particle vacuum state, 
\be
|\Psi\ra=|\psi_n,0_B\ra=\psi_n(q)|0_B\ra,
\ee
where $\hat b_k|0_B\ra=0$ for each $k$. In the first order in the $\hat H_1$, see  (\ref{h1}), the system may decay to another eigenstate $\psi_m(q)$ corresponding to an eigenvalue $E_m$ emitting a single phonon of energy $\epsilon_k$. According to the Fermi golden rule the decay rate reads 
\be
\Gamma = 2\pi\sum_{m}\gamma_m,
\label{G}
\ee
where 
\bea
\gamma_m = |\langle\psi_m,1_k|\hat{H}_1|\psi_n,0_B\rangle|^2 g(\epsilon_k)=|\langle\psi_m|s_k|\psi_n\rangle| g(\epsilon_k). \cr
\label{maleg}
\eea
The sum in (\ref{G}) runs over all eigenstates $\psi_m$ for which we can find such a phonon that the energy conservation $E_n=E_m+\epsilon_k$ is fulfilled. We assume a continuum phonon spectrum (\ref{bogspec}) with the energy gap corresponding to $k=\pi/L$.  The density of states is
\be
g(\epsilon) =
\frac{\epsilon}{\left[2(\epsilon^2+1)\left(\sqrt{\epsilon^2+1}-1\right)\right]^{1/2}}. 
\ee
The lifetime of the Anderson localized states presented in Figs.~\ref{ander}a and \ref{ander}c $\tau=1/\Gamma=8\times10^5$ (17 minutes) and in Figs.~\ref{ander}b and \ref{ander}d $\tau=2.5\times 10^{5}$ (5 minutes)  which means that there is by far enough time to perform experiments until they can decay due to phonon emissions. In Fig.~\ref{gam} we show contributions $\gamma_m$ to the decay rate (\ref{G}) and $\psi_m$ states corresponding to the largest values of $\gamma_m$. Figure~\ref{gam} indicates that the most probable decay leads the system to states localized in the vicinity of the initial localization region. 

Long lifetime of Anderson localized states is very promising from the experimental point of view. Indeed, it means that there is sufficient time to excite a dark soliton in an ultra-cold atomic gas, wait until it localizes in the presence of a weak disorder potential and perform an atom density measurement. If the soliton is Anderson localized distribution of soliton positions collected in many realizations of the experiment \cite{dark} will reveal an exponential profile.

Experiments with dark solitons have been performed in the presence of a harmonic trap \cite{burger1999,andersonPRL01,beckerNP08,stellmerPRL08,wellerPRL08,shomroniNP09}. In such a case, in order to observe the Anderson localization of the solitons, the trap has to be sufficiently shallow. That is, the ground state extension of the soliton position in the harmonic trap without the disorder must be much larger than the localization length predicted in our analysis, i.e. $\frac{1}{\sqrt{|M_q|\omega}}\gg l_{loc}$ where $\omega$ is the harmonic trap frequency.

\section{Conclusions}
\label{concl}

We have considered a dark soliton in dilute ultra-cold atomic gases in the presence of a weak disorder potential. Our consideration is divided into classical and quantum descriptions. The classical approach concerns the analysis of stationary solutions of the Gross-Pitaevskii equation and the effect of deformation of the soliton shape by the disorder. We have employed two methods: the Bogoliubov approach and the expansion of a wavefunction perturbation in eigenmodes of the P\"oschl-Teller potential. These two methods lead to the same results, however, the expansion in the P\"oschl-Teller modes turns out to be more convenient and in particular allows us to obtain a very simple form of the soliton perturbation in terms of an integral kernel. Comparison of the perturbative calculations with the numerical results shows surprisingly good agreement even for the strength of an external potential as great as the chemical potential of the system. If the strength is much smaller than the chemical potential the wavefunction deformation is negligibly small.

The Bogoliubov approach is invaluable in the quantum description where we are interested in many body eigenstates of the system. If the strength of an external potential is much smaller than the chemical potential the dark soliton position may be described by an effective quantum Hamiltonian which is weakly coupled to the quasi-particle subsystem. The effective Hamiltonian turns out to be similar to the corresponding Hamiltonian in the problem of a bright soliton in a weak external potential. Similarly as in the bright soliton case we predict Anderson localization of a dark soliton in the presence of a disorder potential. Because there is a coupling between the soliton position degree of freedom and the quasi-particle subsystem the localized states may decay via phonon emission process. We have investigated lifetimes of the Anderson localized states and it turns out that for typical experimental conditions they exceed condensate lifetimes that make experimental observations of the dark soliton localization realistic.

\section*{Acknowledgments}
This work is supported by the Polish Government within research project
2009-2012 (MM) and by National Science Centre under projects 
DEC-2011/01/N/ST2/00418 (MP) and
DEC-2011/01/B/ST3/00512 (KS).



\end{document}